\documentclass[journal,hidelinks]{IEEEtran}
\usepackage{graphicx,url}
\usepackage[english]{babel}
\addto\captionsenglish{}
\usepackage[utf8]{inputenc}
\usepackage{amsmath}
\usepackage{amssymb}
\usepackage{subfigure}
\usepackage{lipsum}
\usepackage{mathtools}
\usepackage{cuted}
\usepackage{cite}
\usepackage{graphicx}
\usepackage{balance}
\usepackage{tikz}
\usepackage{pgfplots}
\usepackage{tikzscale} 
\pgfplotsset{compat=newest} 
\usepackage{scalefnt}
\usepackage{orcidlink}
\usepackage{dblfloatfix}
\usepackage[acronym,shortcuts]{glossaries}
\usepackage{bm}
\usepackage{amsfonts}
\usepackage{algorithmic}
\usepackage{algorithm}
\usepackage{array}
\usepackage{textcomp}
\usepackage{verbatim}
\usepackage{subcaption}

\newacronym{JCDE}{JCDE}{joint channel and data detection}
\newacronym{ICC}{ICC}{integrated communication and computing}
\newacronym{ISAC}{ISAC}{integrated sensing and communication}
\newacronym{UE}{UE}{user equipment}
\newacronym{ED}{ED}{edge device}
\newacronym{CT}{CT}{channel tracking}
\newacronym{CP}{CP}{channel prediction}
\newacronym{CSI}{CSI}{channel state information}
\newacronym{BER}{BER}{bit error rate}
\newacronym{MSE}{MSE}{mean squared error}
\newacronym{NMSE}{NMSE}{normalized mean squared error}
\newacronym{MMSE}{MMSE}{minimum mean squared error}
\newacronym{UPA}{UPA}{uniform planar array}
\newacronym{ULA}{ULA}{uniform linear array}
\newacronym{SIMO}{SIMO}{single-input multiple-output}
\newacronym{MIMO}{MIMO}{multiple-input multiple-output}
\newacronym{SVD}{SVD}{singular value decomposition}
\newacronym{SIC}{SIC}{soft interference cancellation}
\newacronym{AWGN}{AWGN}{additive white Gaussian noise}
\newacronym{SGA}{SGA}{scalar Gaussian approximation}
\newacronym{VGA}{VGA}{vector Gaussian approximation}
\newacronym{PDF}{PDF}{probability density function}
\newacronym{BS}{BS}{base station}
\newacronym{JCCCT}{JCCCT}{joint communication, computing and channel tracking}
\newacronym{AoA}{AoA}{angle of arrival}
\newacronym{AoD}{AoD}{angle of departure}
\newacronym{AR}{AR}{auto-regressive}
\newacronym{mmWave}{mmWave}{millimeter‑wave}
\newacronym{SotA}{SotA}{state-of-the-art}
\newacronym{AP}{AP}{access point}
\newacronym{BiGaBP}{BiGaBP}{bilinear Gaussian belief propagation}
\newacronym{GaBP}{GaBP}{Gaussian belief propagation}
\newacronym{AirComp}{AirComp}{over-the-air computing}
\newacronym{6G}{6G}{sixth generation}
\newacronym{OFDM}{OFDM}{orthogonal frequency division multiplexing}
\newacronym{QPSK}{QPSK}{quadrature phase shift keying}
\newcommand{\trans}[0]{^{\mathsf{T}}}

\newcommand{\herm}[0]{^{\mathsf{H}}}

\DeclareMathOperator*{\argmin}{argmin}
\DeclareMathOperator*{\diag}{diag}

\begin{document}

\title{Integrated Communication and Computing\\ in Time-Varying mmWave Channels}

\author{Joan Çollaku, Kuranage Roche Rayan Ranasinghe\textsuperscript{\orcidlink{0000-0002-6834-8877}}, \IEEEmembership{Graduate Student Member,~IEEE,} \\
Giuseppe Thadeu Freitas de Abreu\textsuperscript{\orcidlink{0000-0002-5018-8174}}, \IEEEmembership{Senior Member,~IEEE,} and
Takumi Takahashi\textsuperscript{\orcidlink{0000-0002-5141-6247}}, \IEEEmembership{Member,~IEEE,}
% <-this % stops a space
\thanks{J.~Çollaku, ~~K.~R.~R.~Ranasinghe and G.~T.~F.~de~Abreu are with the School of Computer Science and Engineering, Constructor University, Campus Ring 1, 28759 Bremen, Germany (emails: [jcollaku, kranasinghe, gabreu]@constructor.university).}
\thanks{~~T.~Takahashi is with the Graduate School of Engineering, Osaka University, Suita 565-0871, Japan (e-mail: takahashi@comm.eng.osaka-u.ac.jp).}
}


\maketitle

\begin{abstract}
We propose a novel framework for \ac{ICC} transceiver design in time‑varying \ac{mmWave} channels.
In particular, in order to cope with the dynamics of time‑varying \ac{mmWave} channels, the detection of communication symbols and the execution of an \ac{AirComp} operation are performed in parallel with channel tracking, as opposed to existing \ac{SotA} on \ac{ICC} where perfect knowledge of the channel at all time instances is typically assumed.
For clarity of exposition, we consider a \ac{SIMO} uplink scenario where multiple single-antenna \ac{UE} transmit to a \ac{BS} equipped with multiple antennas, such that each \ac{UE}, or \ac{ED}, precodes its own transmit signal, while the \ac{BS}, or \acp{AP}, also performs receive beamforming\footnotemark. 
The proposed transceiver framework then estimates \ac{CSI} and data symbols in parallel, using a \ac{BiGaBP} algorithm for \ac{JCDE}, aided by a \ac{CP} algorithm executed before each estimation window at the \ac{BS}.
The \ac{AirComp} operation is then executed by means of an optimal combination of the residual signal. 
Simulation results demonstrate the effectiveness of the proposed scheme in performing \ac{ICC} in challenging time‑varying \ac{mmWave} channels, with minimal degradation to both communication and computing performance. 

\end{abstract}


\glsresetall

\maketitle

\section{Introduction}

Due to the unrelenting demand for higher data rates and support for an increasingly large number of distinct applications, there is a need for the integration of new functionalities into communication systems.
A prominent example is \ac{ISAC}, which has gained much attention in recent years \cite{Gao2023,Choi2016, LiuJSAC2022, RanasingheTWC2024};
and another is the utilization of the properties of wireless channels to perform \ac{AirComp} \cite{Nazer2007}.
In recognition to this opportunity, multiple works have proposed \ac{AirComp} schemes  \cite{Ando2023,Fang2021,Qin2021,Liu2017,Ranasinghe2025}, more recently giving rise to the notion of \ac{ICC}.

The above works assume, however, perfect knowledge of the channel, which is also a typical assumption made in the design of transceivers for \ac{mmWave} technology.
Indeed, \ac{mmWave} is a key approach to achieve higher rates in future wireless systems \cite{Rangan2014,Wang2018}, but effective beamforming \cite{Adhikary2013, Alkhateeb2014, Liu2019} techniques required to mitigate the high propagation losses and blockage in \ac{mmWave} channels also typically require perfect \ac{CSI}.

Aiming at circumventing this challenge, work has been done to design receivers capable of performing either \ac{CP} \ac{CSI}, or channel tracking, concomitant with the detection of communication symbols.
Among various alternatives, the approach based on \ac{BiGaBP} \cite{Parker2014, Shental2008, Li2024, Takahashi2019} has proved particularly advantageous, due to its excellent trade-off between performance and complexity.

\setcounter{footnote}{1}
\footnotetext{\setlength{\baselineskip}{10pt}Throughout the article the terms \ac{UE} and \ac{BS} will be used interchangeably with \ac{ED} and \ac{AP}, respectively.}

In contribution to both these trends -- of multi-functionality and resilience to time-varying behavior -- we present in this article on a new \ac{BiGaBP}-based framework to perform \ac{JCCCT} over a time-varying \ac{mmWave} channels. 

Receivers designed under the \ac{BiGaBP} framework have been previously employed for joint channel estimation and data detection \cite{Takahashi2019, Takahashi2024} in \ac{MIMO} systems, while a linear \ac{GaBP} receiver has been used for \ac{ICC} in \cite{Ranasinghe2025}. 
Our work follows on these footsteps, integrating to a \ac{BiGaBP} \ac{JCDE} receiver \ac{AirComp}  functionality via a \ac{MMSE} combiner operated over the residual signal \cite{Ranasinghe2025, ranasinghe2025flexible}.

The communication and computing signals are transmitted simultaneously by the \acp{UE}/\acp{ED}s to be received at the \ac{BS}/\ac{AP}, which performs receive beamforming and detection. 
We assume that the channel in our first transmission time instance is known, and the time variation is tracked; then, the \ac{BiGaBP} and \ac{CP} algorithms described in \cite{Takahashi2024} are used to estimate the communication symbols and the channel in parallel, while treating the superimposed computing signal as effective noise.
The \ac{BiGaBP} algorithm performs data estimation by passing messages on a tripartite graph, assuming Gaussian density functions for the interference and noise term in each received signal/channel coefficient, while the \ac{CP} algorithm uses a Kalman filter-like mechanism to provide reliable estimates for the next window of \ac{JCDE}. 

It should be noted that while the \ac{JCDE} \& \ac{CP} algorithm is the same one used in \cite{Takahashi2024}, this work contributes to the SotA by presenting a novel system template that integrates the functionality of AirComp into a high-mobility setting, which has the potential to become an important part of V2X communications. 
Additionally, the proposed framework enables AirComp in mmWave channels.
The proposed system framework can then be used for \ac{ICC} in the \ac{mmWave} band in high-mobility communication scenarios, without having to continually transmit pilot symbols. 

Given this framework, further extensions such as multi-stream computation, or the computation of a wider range of nomographic functions also become possible under the same channel conditions.%
The integration of the \ac{AirComp} operation into the \ac{JCDE} procedure, which is currently performed disjointly from the \ac{JCDE} phase, is also a further avenue to explore in future work.

In this work, we summarize the system model, describe the transmit signal and receiver design, and then evaluate the performance of the scheme in terms of communications \ac{BER} and \ac{NMSE} for channel estimation and \ac{AirComp}.

\textit{Notation:} The following notation is used persistently in the manuscript.
Vectors and matrices are represented by lowercase and uppercase boldface letters, respectively;
$\mathbf{I}_M$ denotes an identity matrix of size $M$ and $\mathbf{1}_M$ denotes a column vector composed of $M$ ones; 
the Euclidean norm and the absolute value of a scalar are respectively given by $\|\cdot\|_2$ and $|\cdot|$;
the transpose and hermitian operations follow the conventional form $(\cdot)\trans$ and $(\cdot)\herm$, respectively;
$\Re{\{\cdot\}}$, $\Im{\{\cdot\}}$ and  $\mathrm{min}(\cdot)$ represents the real part, imaginary part and the minimum operator, respectively.
Finally, $\sim \mathcal{N}(\mu,\sigma^2)$ and $\sim \mathcal{CN}(\mu,\sigma^2)$ respectively denotes the Gaussian and complex Gaussian distribution with mean $\mu$ and variance $\sigma^2$, where $\sim$ denotes ``is distributed as''.

\vspace{-1ex}
\section{System Model}
\subsection{Channel Model}

\begin{figure}[H]
\centering
\includegraphics[width=0.8\columnwidth]{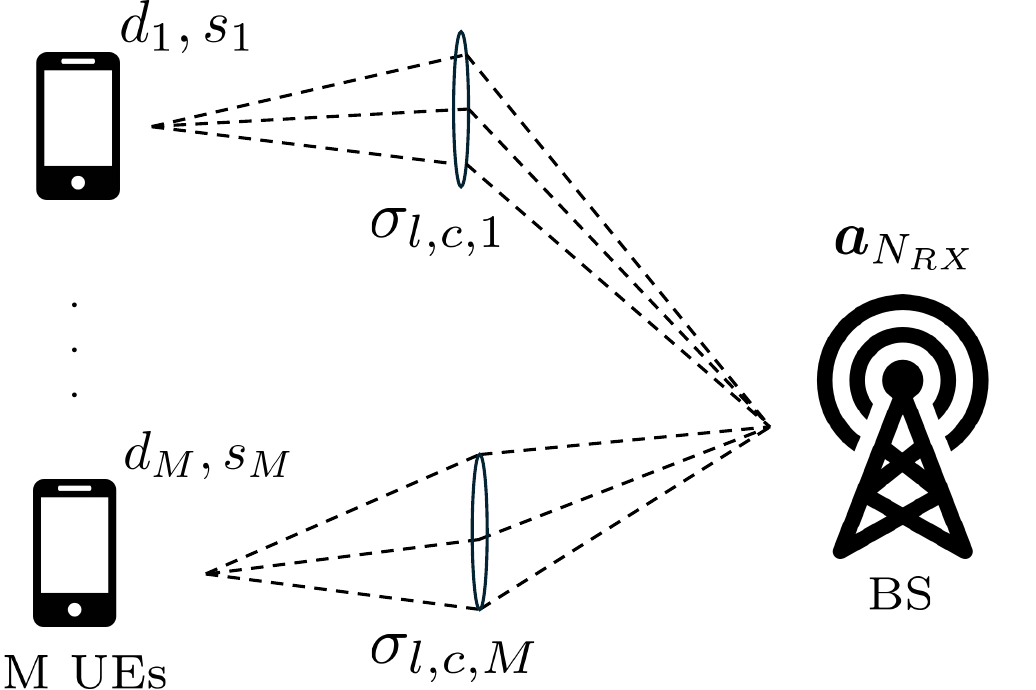}
\vspace{1ex}
\caption{Uplink mmWave SIMO ICC system.}
\label{fig_1}
\vspace{-3ex}
\end{figure}

Consider a multi-user \ac{SIMO} system with $M$ single antenna \acp{UE}/\acp{ED} and a \ac{BS}/\ac{AP} with $N_{RX}$ antennas, which employs receive beamforming as illustrated in Figure \ref{fig_1}. 
Following existing literature \cite{Tsai2018,Xiao2015,Heath2016,Stoica2019}, the channel is modeled using the \ac{mmWave} cluster channel model, with $L$ clusters having $C_l$ rays each, where $l \in \{1, 2, \cdots, L\}$. 
The channel from the $m$-th \ac{UE}/\ac{ED} to the \ac{BS} at a time instance $k$ can be expressed as
\begin{equation}
\label{eq:mmwavechann}
\acute{\boldsymbol{h}}_m[k] = \sum^L_{l=1} \sum^{C_l}_{c=1} \frac{\sigma_{l,c, m}[k]}{\sqrt{LC_l}} \boldsymbol a_{N_{RX}}(\theta_{l,c, m}^{RX}, \phi^{RX}_{l,c, m}) ,
\end{equation}
where $\boldsymbol a_{N_{RX}}(\theta_{l,c, m}^{RX}, \phi^{RX}_{l,c, m})$ is the array response of the \ac{BS}/\ac{AP} receive antennas to the $m$-th \acp{UE}/\acp{ED} signal, $\theta_{RX}$ represents the elevation \ac{AoA}, $\phi_{RX}$ represents the azimuth \ac{AoA} and $\sigma_{l,c, m}$ represents the time-varying small-scale fading coefficient. 

Assuming that the \ac{BS}/\ac{AP} is equipped with a \ac{UPA} with half-wavelength spacing, the array response can be expressed as
\begin{equation}
\label{eq:arrayresponserx}
\boldsymbol{a}_{N}(\theta, \phi) = \boldsymbol{c}_{\sqrt{N_{RX}}}(\sin(\theta) \cos(\phi)) \otimes \boldsymbol{c}_{\sqrt{N_{RX}}}(\cos(\theta)),
\end{equation}
where
\begin{equation}
\label{eq:phasevec}
\boldsymbol{c}_N(\nu) = [1, e^{j\pi \nu},...e^{j\pi(P-1)\nu}].
\end{equation}

For future convenience, a more tractable representation of the channel can be obtained by defining the channel matrix $\acute{\boldsymbol{H}}[k] \triangleq [\acute{\boldsymbol{h}}_1[k], \acute{\boldsymbol{h}}_2[k]...\acute{\boldsymbol{h}}_M[k]] \in \mathbb{C}^{N_{RX} \times M}$, and by extension, the array response matrix $\acute{\boldsymbol{A}_{l,c}} \triangleq [\boldsymbol a_{N_{RX}}(\theta_{l,c, 1}^{RX}, \phi^{RX}_{l,c,1}),...\boldsymbol a_{N_{RX}}(\theta_{l,c, M}^{RX}, \phi^{RX}_{l,c,M})] \in \mathbb{C}^{N_{RX} \times M}$.

The time dependency of the channel can be expressed in our small-scale fading coefficients. 
Given the initial $\sigma_{l,c, m}[0] \sim \mathcal{CN}(0,1)$, future fading coefficients can be expressed using an \ac{AR} model defined by
\begin{equation}
\label{eq:autoreg}
\sigma_{l,c, m}[k] = r\sigma_{l,c, m}[k-1] + \sqrt{1 - r^2}\omega_{l,c, m}[k],
\end{equation}
where $\omega_{l,c, m}[k] \sim \mathcal{CN}(0,1)$ is a random time-varying factor and $r$ is the correlation parameter between time-adjacent \ac{OFDM} symbols.

The procedure for the estimation of $r$ follows consecutively.
The coherence time of the channel is given by
\begin{equation}
\label{eq:coherencetime}
T_c = 0.432 \cdot \frac{v_c}{v}\frac{1}{f_c},
\end{equation}
where $v_c$, $v$ and $f_c$ are the speed of light, the relative velocity between transmitter and receiver and the carrier frequency respectively. 

The symbol duration for a guard interval $N_{G}$ times the total symbol size, an $N_\text{DFT}$ DFT size and a sampling rate $f_s$, can be expressed as
\begin{equation}
\label{eq:symdur}
T_s = N_\text{DFT}\cdot \frac{1 + N_{G}}{f_s}.
\end{equation}

Leveraging the above, the discrete coherence time can be defined as $K_\text{max} \triangleq \lfloor T_c/T_s \rfloor$. 
$r$ can then be approximated by
\begin{equation}
\label{eq:corrcoeff}
r = \exp\left[ \frac{\text{ln}(0.5)}{K_\text{max}}\right].
\end{equation}

\subsection{Signal Model}

Under the assumption of perfect synchronization between users, the received signal in one carrier at a discrete time $k$ is given by
\begin{align}
\label{eq:signalmodel1}
\boldsymbol{y}[k] &= \sum^M_{m=1}\boldsymbol{F}_{RX}\herm \acute{\boldsymbol{h}}_m[k]x_m[k] + \boldsymbol{w}[k] \nonumber \\
&= \sum^M_{m=1}{\boldsymbol{h}}_m[k]x_m[k] + \boldsymbol{w}[k], 
\end{align}
where $\boldsymbol{F}_{RX} \in \mathbb{C}^{N_{RX} \times N}$ is the receive beamformer matrix, $\boldsymbol{h}_m \in \mathbb{C}^{N \times 1}$ is the effective (beam-domain) channel vector of the $m$-th \ac{UE}/\ac{ED} and $\boldsymbol w \sim \mathcal{CN}(0, N_0\boldsymbol I_N)$ is the circularly symmetric \ac{AWGN}.

To incorporate both communications and computing functionality, the transmit signal of each \ac{UE}\ac{ED} is composed of a sum of a communication signal and a computing signal given by
\begin{equation}
\label{eq:iccsignal}
x_m[k] = d_m[k] + \psi_m(s_m[k]),
\end{equation}
where $d_m \in \mathcal{X}$ is a modulated communication symbol from discrete constellation $\mathcal{X}$ (i.e. \ac{QPSK}) and $s_m \in \mathbb{R}$ is the $m$-th computing symbol, preprocessed by a function $\psi_m(\cdot)$.

The target function can be defined as
\begin{equation}
\label{eq:iccfunc}
f(\boldsymbol{s}) = \phi\left( \sum^M_{m=1} \psi_m(s_m) \right).
\end{equation}

For the sake of simplicity, the arithmetic sum is chosen as the target function, i.e. $\phi(\cdot)$ and $\psi_m(\cdot), \forall m$, will be the identity map.

\subsection{Beamforming}

Under the assumption of perfect \ac{CSI}, a popular beamforming scheme in \ac{MIMO} communications uses the \ac{SVD} to diagonalize the channel.
However, under the same assumption, due to the \ac{SIMO} system architecture, data symbols cannot be combined before transmission and therefore, we employ a 'quasi-SVD' combiner. 
The construction of the receive beamformer will be the same as in the \ac{MIMO} case, since the same receiver architecture is being used.
At the transmitter, since we utilize single-antenna \acp{UE}/\acp{ED}, the beamformer is limited. 
Let us start with the fact that the time zero (known) channel can be decomposed as
\begin{equation}
\label{eq:svd}
\acute{\boldsymbol H}[0] = \boldsymbol{U\Sigma V}\herm.
\end{equation}

The combiner can then be constructed as
\begin{equation}
\label{eq:rxbf}
\boldsymbol{F}_{RX} = [\boldsymbol{U}]_{:, 1:N},
\end{equation}
where $N \leq N_{RX}$. 
If $N$ is significantly larger than the number of clusters in our \ac{mmWave} channel, we can fully exploit the channel's degrees of freedom. 

This approach will not result in a fully diagonal effective channel, but instead, an upper-triangular channel matrix, which does not cause a significant hindrance to the performance of our detection scheme. 
The combiner is kept constant as the channel varies, so the effective channel can be formulated as
\begin{equation}
\label{eq:bfchann}
\boldsymbol{H}[k] =  \boldsymbol F_{RX}\herm \acute{ \boldsymbol H}[k] \in \mathbb{C}^{N \times M}.
\end{equation}

For convenience, the effective channel can also be expressed as
\begin{equation}
\label{eq:effchann}
\boldsymbol{H}[k] = \sum^L_{l=1} \sum^{C_l}_{c=1} \frac{1}{\sqrt{LC_l}} \boldsymbol{A}_{l,c}\boldsymbol\sigma_{l,c}[k] ,
\end{equation}
where $\boldsymbol{A}_{l,c} = \boldsymbol F_{RX}\herm \acute{ \boldsymbol{A}_{l,c}} \in \mathbb{C}^{N \times M}$ is the beam-domain array response matrix, and $\boldsymbol \sigma_{l, c} \triangleq \diag(\sigma_{l,c, 1}[k], \dots \sigma_{l,c, M}[k])$ is the small-scale fading coefficient matrix of the $c$-th ray of the $l$-th cluster at time $k$.

\section{Proposed Joint Communication, Computing and Channel Tracking Receiver}

In this section, the framework for joint channel estimation/tracking and data detection proposed in \cite{Takahashi2019} is augmented with the integration of \ac{AirComp}, where we compute a target function with the computing symbols sent by each user as input.
$E_d$ represents the power allocated to the communications symbols and $E_c$ is the power is allocated to the computing symbols such that $E_d + E_c = 1$.
The communication symbols are modulated with \ac{QPSK} and the computing symbols are normally distributed with zero mean; $s_k \sim \mathcal{N}(0, E_c)$.
The soft replicas of the channel are defined as $\hat{h}_{nm, k}$, $\hat{\boldsymbol{h}}_{m, k}$, $\hat{\boldsymbol{H}_k}$ respectively for each channel coefficient, channel vector, and channel matrix respectively, with the respective MSEs being defined as $\hat \psi^h_{nm, k}$, $\hat {\boldsymbol {\varPsi}}^h_{m, k}$ \& $\hat {\boldsymbol {\varPsi}}^h_{k}$.

The second-order statistics of the channel, assuming knowledge of a past channel, which are going to be useful in \ac{JCDE} and \ac{CP} are additionally calculated in \cite{Takahashi2024}, with their closed forms given by

\begin{subequations}
\label{eq:channelvar}
\begin{eqnarray}
\omega_{nm,k'} \hspace{-4ex}&&= \mathbb{E}\left[ |h_{nm}[k] - r^{k'}h_{nm}[k - k']|^2 \right] \\ 
&&= \frac{1 - r^{2k'}}{L}\sum^L_{l = 1} \sum^{C_l}_{c = 1} \frac{|[\boldsymbol{A}_{l, c}]_{n, m}|^2}{C_l} = \frac{1 - r^{2k'}}{L} \theta_{nm}, \nonumber
\end{eqnarray}   
\begin{eqnarray}
\!\!\boldsymbol \Omega_{m,k'}  
\hspace{-4ex}&&\!=\! \mathbb{E}\left[ (\boldsymbol{h}_m[k]\! -\! r^{k'}\boldsymbol{h}_m[k\!-\!k'])(\boldsymbol{h}_m[k]\!- \!r^{k'}\boldsymbol{h}_m[k\!-\!k'])\herm \right]\!\!\! \nonumber \\ 
&&= \frac{1\!-\!r^{2k'}}{L}\sum^L_{l = 1} \sum^{C_l}_{c = 1} \frac{[\boldsymbol{A}_{l, c}]_{:, m} [\boldsymbol A_{l, c}]_{:, m}\herm}{C_l},
\end{eqnarray}   
\begin{eqnarray}
\boldsymbol \Omega_{k'} \hspace{-4ex}&&=\mathbb{E}_{\boldsymbol{H}[k']} \left[\boldsymbol{H}[k] \boldsymbol{H}[k]\herm |\boldsymbol{H}[k - k'] \right] \\ 
&&= \frac{1 - r^{2k'}}{L}\sum^L_{l = 1} \sum^{C_l}_{c = 1} \frac{\boldsymbol A_{l, c} \boldsymbol A_{l, c}\herm}{C_l} = \frac{1 - r^{2k'}}{L} \boldsymbol \Theta,\nonumber
\end{eqnarray}
where $\theta_{nm}$ and $\boldsymbol \Theta$ signify the true variance of the channel coefficients and the covariance matrix of the channel, respectively.
\end{subequations}

For a considered set of time indices $\mathcal K$, the algorithm is performed in time windows $\mathcal K_\tau \subset \mathcal{K}$, which, due to an unchanging relative velocity assumption, can be constructed as having a constant length (determined by parameter $W$) and a constant number of new samples with respect to the previous window (determined by parameter $D$).
The $\tau$-th window can be defined as
\begin{equation}
\label{eq:window}
\mathcal{K}_\tau = \{\,k \in \mathcal{K} \mid (\tau - D)W \le k \le \tau W - 1\},
\end{equation}
which consequently leads to $\tau_{max} = K/W + D - 1$ windows. 

At each window, \ac{CP} is carried out to provide better starting estimates, which are then utilized in the message passing algorithm for the data detection stage and then as a starting point for channel estimation.

\subsection{Channel Prediction}

The \ac{CP} algorithm works by calculating the conditional expectation of the new channels given the known channels within any given window.
Given knowledge of the channel at $k = 0$, the expectation can be expressed as
\begin{equation}
\label{eq:channmean}
\mathbb{E}[\boldsymbol{H}[k]|\boldsymbol{H}[0]] = r^k \boldsymbol{H}[0],
\end{equation}
with the the conditional variances being $\omega_{nm, k}$, $\boldsymbol \Omega_{m, k}$, and  $\boldsymbol \Omega_{k}$ $\forall (n,m)$ for each channel coefficient, vector and matrix respectively. 

In windows where $\tau \neq 1$, and the known channel is not available, the conditional expectation of the channel given the previously estimated channels that are still within the \ac{CP} window is found. 
Note that the sub-optimality of the proposed \ac{JCDE} algorithm results in the reliability of our channel estimates varying randomly with time. 
To alleviate this effect, the prior estimate that builds on the predictions is going to be the estimate $k_\tau$ with the minimum \ac{MSE} among the previously estimated channels in the current \ac{CP} window, expressed as
\begin{equation}
\label{eq:smallk}
k_{\tau}
= \argmin_{k \in \mathcal{K}_{\tau} \setminus \mathcal{K}_{\tau}^{+}}
\sum_{n=1}^{N} \sum_{m=1}^{M}
\hat{\psi}^{\mathrm{h}}_{k,nm},
\end{equation}
where $\hat{\psi}_{nm, k}$ denotes the \ac{MSE} of a channel coefficient \ac{JCDE} estimate at time index $k$ and $\mathcal K_\tau^+ = \mathcal K_\tau \backslash(\mathcal K_\tau \cap \mathcal K_{\tau-1})$ i.e. the set of new additions to the current time window. 

\begin{subequations}
\label{eq:CP}
After obtaining this \ac{MSE} channel estimate, the predictions for the channel and the \ac{MSE} of the channel estimates are updated as
\begin{eqnarray}
&\hat{\boldsymbol{H}}_k = \mathbb{E}\left[\boldsymbol{H}[k]|\hat {\boldsymbol{H}}_{k_\tau}\right] = r^{k - k_\tau}\hat {\boldsymbol{H}}_{k_\tau},&\\
&\hat{\psi}^h_{nm, k} = \omega_{nm, k-k_\tau} + r^{2(k-k_\tau)}\hat{\psi}^h_{nm, k_{\tau}},&\\
& \hat{\boldsymbol \Psi}^h_{m, k} = \boldsymbol \Omega_{m, k-k_\tau} + r^{2(k-k_\tau)}\hat{\boldsymbol \Psi}^h_{m, k_{\tau}},&\\
& \hat{\boldsymbol \Psi}^h_{k} = \boldsymbol \Omega_{k-k_\tau} + r^{2(k-k_\tau)}\hat{\boldsymbol \Psi}^h_{k_{\tau}}.&
\end{eqnarray}
\end{subequations}

Note that these equations are only used if $k > k_\tau$. 
If not, the estimates are not updated.

\subsection{Data Detection}

To initialize the algorithm, all communication symbol soft replicas $\hat{d}_{m,k}$ are set to 0 and their \acp{MSE} $\hat{\psi}_{m,k}^d = \mathbb{E}[|x_m[k] - \hat{x}_{m,k}|^2]$ are set to 1. 
During detection, the computing symbols are treated as effective noise, given by
\begin{equation}
\label{eq:signalmodel}
\boldsymbol y[k] = \boldsymbol{H} [k]\boldsymbol d[k] + (\boldsymbol{H} [k]\boldsymbol{s}[k] + \boldsymbol{w}[k]) = \boldsymbol{H} [k]\boldsymbol d[k] + \tilde{\boldsymbol{w}}[k],
\end{equation}
where $\boldsymbol{w} \sim \mathcal{CN}(0, \tilde{N}_0)$ is the effective noise with power $ \tilde{N}_0 = N_0 + E_c$.

% \vspace{-3ex}
\subsubsection{Factor Nodes}

The \ac{SIC} procedure is performed on the factor nodes, and can be expressed as
\begin{align}
\label{eq:ddsic}
&\tilde{\boldsymbol y}_{m, k} = \boldsymbol{y}[k] - \sum_{i \neq m}^M \hat{\boldsymbol h}_{k, i} \hat{d}_{i, k} = \hat{\boldsymbol{h}}_{m, k}d_{m}[k] \\
&\hspace{6ex}+ \underbrace{ \tilde{\boldsymbol{h}}_{m, k}d_{m}[k] + \sum^M_{i \neq m}(\boldsymbol{h} _i[k]d_i[k] - \hat{\boldsymbol{h}}_{i, k}\hat{d}_{m, k}) + \tilde{\boldsymbol{z}}[k]}_\text{Residual Interference and Noise $(\boldsymbol{\epsilon}^d_{k,m})$}. \nonumber
\end{align}

Using the \ac{VGA}, the \ac{PDF} of $\boldsymbol{\epsilon}^d_{k,m}$ can be approximated by a Gaussian mixture model $p(\tilde{\boldsymbol y}_{m, k}|d_m[k])$ with mean vector $\hat{\boldsymbol h}_{m,k}d_m[k]$ and covariance matrix $\boldsymbol{\Xi}_{m, k}$, which is given by
\begin{align}
\label{eq:xik}
\boldsymbol{\Xi}_{m, k} &= \mathbb{E}[\boldsymbol{\epsilon}_{k,m}^d]\\
&= \underbrace{\sum^M_{i = 1}\hat{\boldsymbol{h}}_{i, k}\hat{\boldsymbol{h}}_{i, k}\herm \hat{\psi}^d_{ik} + \tilde{N_0}\boldsymbol{I}_N + \hat{\boldsymbol{\Psi}}^h_k}_{\boldsymbol{\Xi_k}} - \hat{\boldsymbol{h}}_{i, k}\hat{\boldsymbol{h}}_{i, k}\herm\hat{\psi}^d_{mk}, \nonumber
\end{align}

\vspace{-2ex}
\subsubsection{Variable Nodes}

Beliefs are combined across receive antennas in the variable nodes, leading to a joint belief. 
The PDF of these combined beliefs beliefs can be expressed by reformulating the vector distribution as a scalar, as
\begin{equation}
\label{eq:pdfdata}
p(\tilde{\boldsymbol y}_{m, k}|d_m[k]) = \exp \left[ 
\frac{|d_m[k] - \bar{d}_{m,k}|^2}{\bar{\psi}_{m, k}^d}
\right],
\end{equation}
with
\begin{subequations}
\label{eq:dd_bar}
\begin{eqnarray}
&\bar{d}_{m, k} = \dfrac{1}{\eta_{m, k}}\hat{\boldsymbol{h}}_{m, k}\herm \boldsymbol{\Xi}_m \tilde{\boldsymbol y}_{m, k},&\\
%
%\label{eq:dd_bfvar}
&\bar{\psi}_{m, k}^d = \dfrac{1 - \eta_{m, k}\hat{\psi}^d_{m, k}}{\eta_{m, k}},&
\end{eqnarray}
\end{subequations}
where $\eta_{m, k} = \hat{\boldsymbol{h}}\herm_{m, k} \boldsymbol{\Xi}_m \hat{\boldsymbol{h}}_{m, k}$ is a term derived from the matrix inversion lemma, to eliminate the dependency of the used $\boldsymbol \Xi$ matrix on $m$.

% \vspace{-3ex}
\subsubsection{Denoising and damping}

\begin{subequations}
\label{eq:dddenoise}
The Bayes-optimal denoiser for \ac{QPSK} symbols is used to obtain new estimates for the soft replicas, which are then damped to ease convergence. 
The denoiser is given by
\begin{eqnarray}
&\!\!\!\hat{d}_{m, k}' \!=\! c_d\Big(\!
\tanh \big( 2c_d
\frac{{\Re}(\bar{d}_{m, k})}{\bar{\psi}^d_{m, k}}
\big) \!+ \!
j\tanh \big( 2c_d
\frac{{\Im}(\bar{d}_{m, k})}{\bar{\psi}^d_{m, k}}
\big)\!
\Big),&\;\;\;\;\\
&\hat{\psi'}_{m, k}^d = 1 - \left \lvert \hat{d}_{m, k}\right \rvert ^2,&
\end{eqnarray}
where $c_d = \sqrt{E_d/2}$ is the real and imaginary part of the QPSK symbols transmitted. 
\end{subequations}

\begin{subequations}
\label{eq:dddamp}
Then, damping with a coefficient $ 0\leq \beta \leq 1$ yields
\begin{eqnarray}
&\hat{d}_{m, k} = \beta\hat{d}_{m, k}' + (1-\beta)\hat{d}_{m, k},
\\
&\hat{\psi'}_{m, k}^d = \beta\hat{\psi'}_{m, k}^d
+ (1-\beta)\hat{\psi}_{m, k}^d.
\end{eqnarray}
\end{subequations}

\subsection{Channel Estimation}

Channel coefficient beliefs have to be propagated in the time dimension for this phase, to make full use of our receive diversity. Due to this, a neighborhood is defined for the combining of beliefs in each time instance based on a window length parameter $G$, so as to avoid error propagation. 
At the $\tau$-th iteration, the set around index $k$ is defined as
\begin{equation}
\label{eq:sset}
\mathcal{S}_{k, \tau} = 
\left\{
s \in \breve {\mathcal{K}}_\tau\backslash \{k\} \middle \vert k-\frac{G}{2} \leq s\leq k + \frac{G}{2}
\right\},
\end{equation}
where $\breve {\mathcal{K}}_\tau = \bigcup^\tau_{t = 1} \mathcal{K}_t$.

\begin{algorithm}[H]
\caption{Proposed Joint Communication, Computing and Channel Estimation Algorithm}
\label{alg:jccct}
\begin{algorithmic}[1]
\REQUIRE $\boldsymbol{y}[k], \forall k$, $\boldsymbol A_{l,c} \forall l, c$ , $\boldsymbol H[0]$, r, $\tilde{N}_0$ , $W$, $G$, $D$, $t_\text{max}$    % now prints “Input: X, Y”
\ENSURE $\hat{\boldsymbol d}[k]$, $\hat{\boldsymbol H}[k]$,  $\hat f(\boldsymbol s[k])$, $\forall k$ \\        % now prints “Output: Z”
PREPROCESSING\\
\STATE Obtain $\mathcal{K}_\tau$ using (\ref{eq:window}), generate $\mathcal{K}_\tau^+$ and $\breve{\mathcal{K}}_\tau$
\STATE Obtain $\mathcal{S}_{k,\tau}$ using (\ref{eq:sset})
\STATE Initialize $\hat{d}_{m,k} = 0, \psi^d_{m, k} = 1$, $\forall (m, k)$
\STATE Obtain $\theta_{nm}$, $\boldsymbol{\Theta}_{m}$, $\forall (n, m)$, using equations (\ref{eq:channelvar}) respectively.
\STATE Obtain $\omega_{nm, k}$, $\boldsymbol{\Omega}_{m, k}$, $\boldsymbol{\Omega}_{k}$, $\forall (n, m, k)$ using (\ref{eq:channelvar}) respectively.\\
CHANNEL PREDICTION\\
\FOR{$\tau = 1$ \TO $\tau_{max}$}
\IF{$\tau = 1$, $\forall k \in \mathcal{K}_1$}
\STATE $\hat {\boldsymbol{H}}_k = r^k\boldsymbol{H}[0], \:\hat{ \boldsymbol \Psi}_{k}\herm$ = $\boldsymbol{\Omega}_k$
\STATE $\forall m$ $\hat{ \boldsymbol \Psi}_{k, m}^h = \boldsymbol{\Omega}_{k,m}$
\STATE $\forall (n, m)$ $\hat{ \psi}_{nm, k}^h =\omega_{nm, k}$, $\forall $
\ELSE
\STATE $(\forall k \in \mathcal{K}_\tau, n, m)$ Obtain $\hat {\boldsymbol{H}}_k$, $\hat{ \boldsymbol \Psi}_{k}\herm$, $\hat{ \boldsymbol \Psi}_{k, m}\herm$, $\hat{ \psi}_{nm, k}^h$ using equations (\ref{eq:CP}).
\ENDIF\\
JCDE\\

\FOR{t = 1 \TO $t_\text{max}$, $\forall k \in \mathcal{K}_\tau$}

\STATE $\forall m$, obtain $\tilde{\boldsymbol y}_{m, k}$ using equation (\ref{eq:ddsic}).
\STATE Obtain $\boldsymbol \Xi_{k}$ using equation (\ref{eq:xik}).
\STATE $\forall m$, $\eta_{m, k} = \hat{\boldsymbol{h}}\herm_{m, k} \boldsymbol{\Xi}_m \hat{\boldsymbol{h}}_{m, k}$
\STATE $\forall m$, obtain $\bar{d}_{m, k} \;and\; \bar{\psi}_{m, k}^d$ using equations (\ref{eq:dd_bar})
\STATE $\forall m$, obtain $\hat{d}_{m, k}'$ and $\hat{\psi'}_{m, k}^d$ using equations (\ref{eq:dddenoise})
\STATE $\forall m$, update $\hat{d}_{m, k}$ and $\hat{\psi}_{m, k}^d$ using (\ref{eq:dddamp}).
\STATE $\forall (m, n)$, obtain $\tilde{y}_{nm, k}$ using equation (\ref{eq:ceic})
\STATE $\forall (m, n)$, obtain $\nu_{nm, k}$ using equation (\ref{eq:nu})
\STATE $\forall (s \in \mathcal S_{k, \tau}, m, n)$, obtain $\nu_{nm, s \rightarrow k}$ using (\ref{eq:nuarr}). \\$\forall (m, n)$
\IF{$t = t_\text{max}$}
\STATE $\bar{\psi}_{nm, k}^h = \left(
\sum_{s \in \mathcal{S}_{k, \tau} \cup \{k\}} \frac{|\hat{d}_{m, s}|^2}{\nu_{s \rightarrow k, nm}}
\right)^{-1}$
\STATE $\bar{h}_{nm, k} =
\bar{\psi}_{nm, k}^h\sum_{s \in \mathcal{S}_{k, \tau} \cup \{k\}}\frac{\hat{d}^*_{nm, s}r^{k-s}\tilde{y}_{nm, s}}{\nu_{s \rightarrow k, nm}}$

\ELSE
\STATE $\forall (m, n)$, obtain $\bar{h}_{nm, k}$ and $\bar{\psi}_{nm, k}^h$ using equations (\ref{eq:cebelief}).
\ENDIF
\STATE $\forall m$, obtain $\hat{ \boldsymbol{h}}_{m, k}'$ and $\hat{\boldsymbol \Psi'}_{k, m}^h$ using equations (\ref{eq:cedenoise}).
\STATE $\forall m$, update $\hat{ \boldsymbol{h}}_{m, k}$ and $\hat{\boldsymbol \Psi}_{k, m}^h$ using (\ref{eq:cedamp})
\ENDFOR
\ENDFOR
\STATE $\forall (m, k)$, $\hat{d}_m[k] = \argmin_{d \in \mathcal{X}}\left|d - \hat{d}_{m,k}\right|$
\STATE $\forall k$, $\hat{\boldsymbol{H}}[k] = \hat{\boldsymbol{H}}_k$
\\
AIRCOMP\\
\STATE $\forall k \in \mathcal{K}$, Obtain $\boldsymbol u_k$ using equation (\ref{eq:aircomp}).
\STATE $\forall k \in \mathcal{K}$, Obtain $\hat{f}(\boldsymbol s[k])$ using equation (\ref{eq:combiner}).
\end{algorithmic}
\end{algorithm}

\subsubsection{Factor Nodes} 

The \ac{SIC} is carried out as
\begin{equation}
\label{eq:ceic}
\tilde{y}_{nm, k} = y_n[k] - \sum^M_{i \neq m} \hat{h}_{ni, k}\hat{d}_{i, k}.
\end{equation}

Assuming that the effective noise component can be approximated via the \ac{SGA} to model errors due to \ac{AWGN}, the computing signal, interference and channel aging, we can extract the following \ac{PDF} for the aged \ac{SIC} terms $r^{k-s}\tilde{y}_{nm, s}$ as
\begin{equation}
\label{eq:pdfchann}
\begin{aligned}
p(r^{k-s}\tilde{y}_{nm, s} | h_{nm}[k]) 
\propto \exp \left[
\frac{|r^{k-s}\tilde{y}_{nm, s} - h_{nm}[k]\hat{d}_{mk}|^2}{\nu_{s\rightarrow k, nm}}
\right],
\end{aligned}
\end{equation}
where
\begin{align}
\label{eq:nuarr}
\nu_{s\rightarrow k, nm} =&\\
&\hspace{-6ex}
\begin{cases}
\omega_{s-k, nm}\left|\hat{d}_{m, s}\right|^2 + r^{2(k-s)}\nu_{s, nm} &  k > s \in \mathcal{S}_{k, \tau},\\
r^{2(k-s)}\left(
\omega_{s-k, nm}\left|\hat{d}_{m, s}\right|^2 + \nu_{s, nm}
\right) &   k < s \in \mathcal{S}_{k, \tau},
\end{cases}\nonumber
\end{align}
and $\nu_{s, nm}$ is defined as
\begin{align}
\label{eq:nu}
\nu_{s, nm} &= \sum^M_{i \neq m} \left\{
\left| \hat{h}_{ni, k} \right|^2 \hat{\psi}^d_{i, k} + \left( |\hat{d}_{m, i}|^2 + \hat{\psi}^d_{i, k} \right)\psi^h_{ni, k}
\right\} \nonumber \\
&+ \theta_{nm}\hat{\psi}^d_{m, k}+ \tilde{N}_0.
\end{align}

The dual formula for the interference-cancelled term's squared error (\ref{eq:nuarr}) lends itself to the fact that we are propagating beliefs in the time dimension; beliefs propagated from the past and from the future have different statistical properties, as the error due to channel aging is scaled differently in the calculation of the effective noise component.
Since the aforementioned error is zero-mean, however, this does not affect the \ac{SIC} means.
The $\theta_{nm}$ term is used as a substitute to the true channel gain $\left|h_{nm}[s]\right|^2$ at time instances. 

% \vspace{-2ex}
\subsubsection{Variable Nodes}

At this stage, assuming the \ac{SGA}, the effective noise \ac{PDF}s of each belief are multiplied to get the extrinsic beliefs
\begin{equation}
\!\!\!\prod_{s \in \mathcal{S}_{k, \tau}}\!\! p(r^{k-s}\tilde{y}_{nm, s} | h_{nm}[k]) \propto \exp\left[
\frac{h_{nm}[k]\! -\! \bar{h}_{nm, k}}{\bar{\psi}_{nm, k}^h}
\right]\!,
\end{equation}
where
\begin{subequations}
\label{eq:cebelief}
\begin{eqnarray}
&\bar{h}_{nm, k} =
\bar{\psi}_{nm, k}^h\sum_{s \in \mathcal{S}_{k, \tau}}\dfrac{\hat{d}^*_{nm, s}r^{k-s}\tilde{y}_{nm, s}}{\nu_{s \rightarrow k, nm}},\\
&\bar{\psi}_{nm, k}^h = \left(
\sum_{s \in \mathcal{S}_{k, \tau}} \dfrac{|\hat{d}_{m, s}|^2}{\nu_{s \rightarrow k, nm}}
\right)^{-1}\!\!\!\!\!\!.&
\end{eqnarray}
\end{subequations}

The extrinsic beliefs are generated, since their combining over the time dimension can alleviate error feedback from the \ac{SIC} stage.

% \vspace{-2ex}
\subsubsection{Denoising and damping}

A Gaussian denoiser, obtained as an expectation of the channel vectors, conditioned  on the extrinsic belief \ac{PDF}s, is used for the soft replica updating for both the soft replicas and their \ac{MSE}. 
First, let us define $\bar{\boldsymbol h}_{m, k} = [\bar{h}_{1m, k}, \dots \bar{h}_{Nm, k}]$, $\bar{\boldsymbol \Psi}_{k, m}^h = \diag[\bar{\psi}_{k, 1m}^h,\dots \bar{\psi}^h_{k, Nm}]$ and the sum of the past and new covariance matrices $\boldsymbol \Lambda_{m, k} = \boldsymbol{\Omega}_{m, k} + \bar{\boldsymbol \Psi}_{k, m}^h$. 
Then, the new replicas can be generated as
\begin{subequations}
\label{eq:cedenoise}
\begin{eqnarray}
&\hat{ \boldsymbol{h}}_{m, k}' = \boldsymbol{\Omega}_{m, k}\boldsymbol \Lambda^{-1}_{m, k}\bar{\boldsymbol h}_{m, k} + r^k\bar{\boldsymbol \Psi}_{k, m}^h\boldsymbol \Lambda^{-1}_{m, k}\boldsymbol{h}_m[k],\\
&\hat{\boldsymbol \Psi'}_{k, m}^h =  \boldsymbol{\Omega}_{m, k}\boldsymbol \Lambda^{-1}_{m, k}\bar{\boldsymbol \Psi}_{k, m}^h.
\end{eqnarray}
\end{subequations}

\begin{subequations}
\label{eq:cedamp}
Finally, damping is performed with the same coefficient $\beta$, yielding
\begin{eqnarray}
&\hat{ \boldsymbol{h}}_{m, k}' = \beta\hat{ \boldsymbol{h}}_{m, k}' + (1 - \beta)\hat{ \boldsymbol{h}}_{m, k},\\
&\hat{\boldsymbol \Psi'}_{k, m}^h = \beta \hat{\boldsymbol \Psi'}_{k, m}^h  + (1 - \beta)\hat{\boldsymbol \Psi}_{k, m}^h.
\end{eqnarray}
\end{subequations}

\subsection{AirComp}

After successful detection for the communication symbols and channel estimation, we perform \ac{AirComp} to obtain an estimate of the target function at time $k$ using an \ac{MMSE} combiner $\boldsymbol u_k \in \mathbb{C}^{N \times 1}$ on the channel residual as
\begin{equation}
\label{eq:aircomp}
\hat{f}(\boldsymbol{s}[k]) = \boldsymbol{u}_k\herm(\boldsymbol{y}[k] - \hat{\boldsymbol{H}}[k]\hat{\boldsymbol{d}}[k]).
\end{equation}

The \ac{MMSE} combiner is computed as the solution to the problem
\begin{align}
\label{eq:combinerdef}
\boldsymbol{u}_k & = \argmin_{\boldsymbol{u}_k \in \mathbb{C}^{N \times 1}} \left|\left| f(\boldsymbol s) -  \hat{f}(\boldsymbol s) \right|\right|^2_2\\
& = \argmin_{\boldsymbol{u}_k \in \mathbb{C}^{N \times 1}} \left|\left| \boldsymbol{1}_M\trans \boldsymbol{s}[k] - \boldsymbol{u}_k\herm(\boldsymbol{y}[k] - \hat{\boldsymbol{H}}[k]\hat{\boldsymbol{d}}[k])  \right|\right|^2_2. \nonumber
\end{align}

The combiner for time index $k$, which was computed without considering channel estimation error, is given by
\begin{equation}
\label{eq:combiner}
\boldsymbol{u}_k = (\hat{\boldsymbol{H}}[k] (\boldsymbol {\xi}_k + E_c \cdot \boldsymbol{I_M}) \hat{\boldsymbol{H}}[k]\herm + N_0\boldsymbol{I}_N)^{-1}E_c \hat{\boldsymbol{H}}[k]\cdot \boldsymbol{1}_M,
\end{equation}
where $\boldsymbol {\xi}_k \in \mathbb{C}^{M \times M}$ is the communication data estimation error covariance matrix, which is defined as $\diag[\psi^d_{1, k}, \dots \psi^d_{M, k}]$.

% \section{Algorithm}
The algorithmic description of the proposed receiver is summarized  in Algorithm \ref{alg:jccct}. In addition to receive signals $\boldsymbol{y}[k], \forall k$, the algorithm takes as input the array response matrices $\boldsymbol A_{l,c}$, the known channel $\boldsymbol H[0]$, the channel correlation coefficient $r$, the effective noise spectral density $\tilde{N}_0$ and the algorithm design parameters $(W, D, G)$, as well as the number of JCDE iterations $t_\text{max}$. 
The output of the algorithm will be the detected communication symbols $\hat{\boldsymbol d}[k]$, the estimated channel $\hat{\boldsymbol H}[k]$ and the target function estimated output $\hat f(\boldsymbol s[k])$. 
The complexity of the algorithm is dominated by matrix inversions in the data detection stage ($\boldsymbol \Xi_k$), in the channel estimation stage ($\boldsymbol{\Lambda}_{m, k}$) and additionally in the \ac{AirComp} stage for the combiner computation.

\section{Performance Analysis}

A system with $N_{RX} = 16$ receive antennas, $P = 2$ receive antennas, $M=2$ users, and $N = 8$ over $K = 128$ transmissions over time was simulated in MATLAB. 
The system was evaluated across different relative velocities (10-40 km/h) between transmitter and receiver, assuming the different transmitting \acp{UE}/\acp{ED} are moving at the same velocity. 
The system assumes an \ac{OFDM} carrier frequency of $f_c = 60$ GHz with an \ac{OFDM} sampling frequency of $f_s = 2.64$GHz. 
The DFT size was set to 512 and the guard interval to $1/4$-th of it \cite{Cordeiro2010}. 
The \ac{mmWave} channel is modeled as having $L = 4$ clusters with $C_l = 15$ rays each.
The algorithm parameters were set to $(W, D, G) = (8, 3, 6)$, $t_\text{max} = 8$ and $\beta = 0.5$. 
The transmit power was set to $E_d = 0.99$ for communications and $E_c = 0.01$ for communications. 

\begin{figure}[H]
\subfigure[{\footnotesize BER performance of the system for different relative velocities.}]%
{\includegraphics[width=0.975\columnwidth]{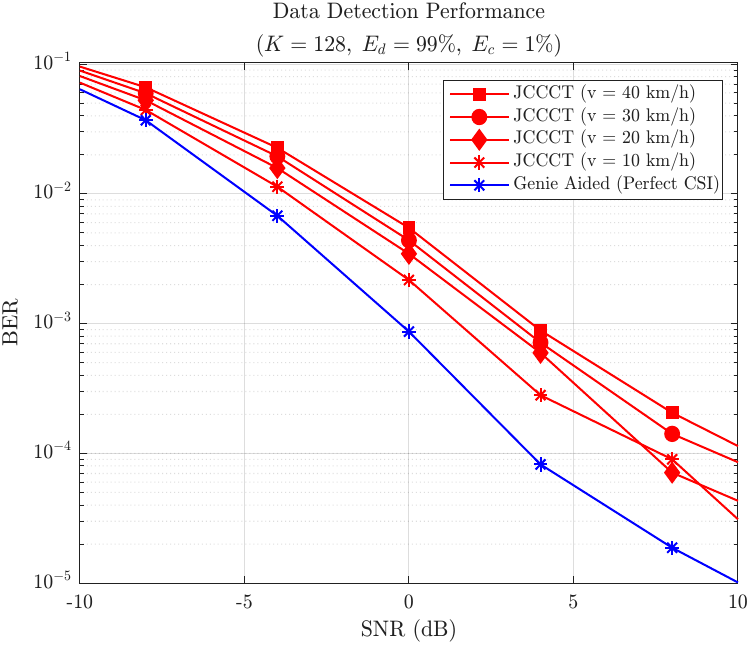}
\label{BER}}\\
\subfigure[{\footnotesize Channel estimation performance for different relative velocities.}]%
{\includegraphics[width=0.975\columnwidth]{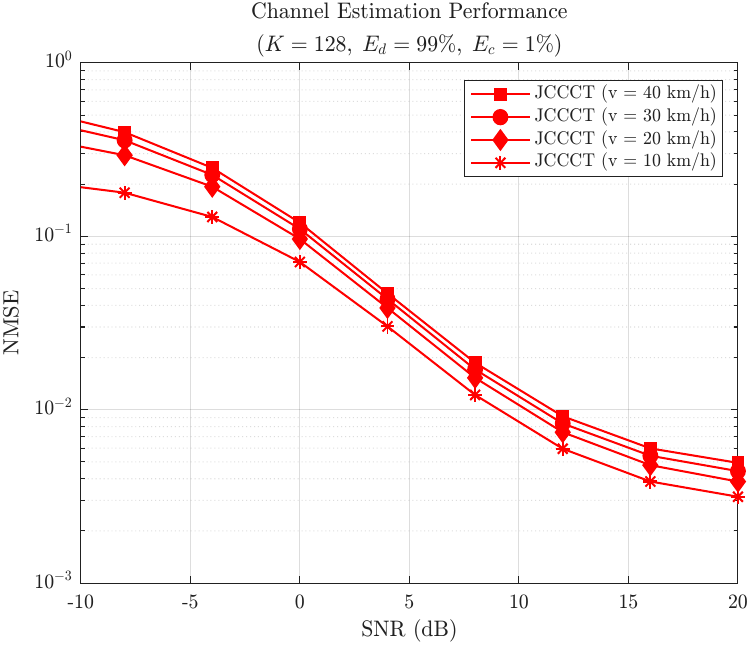}
\label{CE}}
\subfigure[{\footnotesize AirComp performance compared to bounding references.}]%
{\includegraphics[width=0.975\columnwidth]{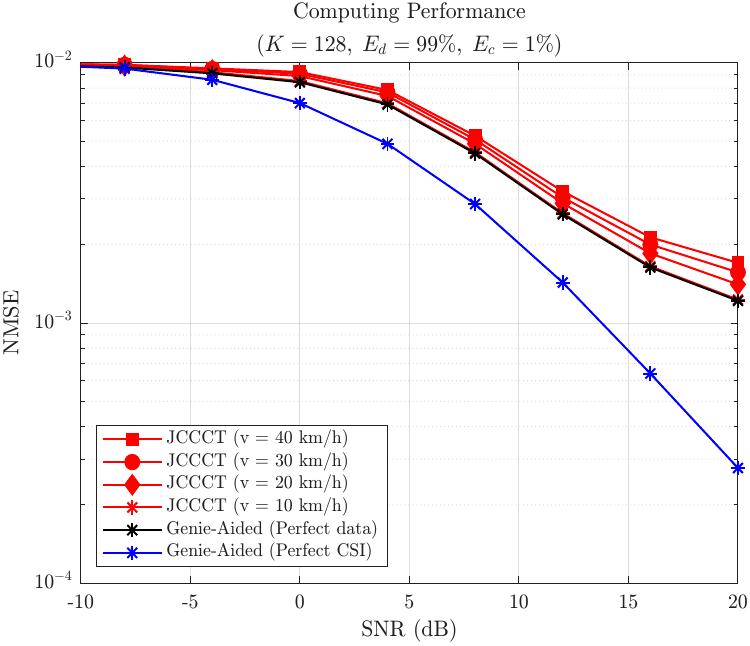}
\label{AirComp2}}
\end{figure}

\newpage
The BER was plotted for the communication symbols (Figure \ref{BER}), and the NMSE was plotted for the channel estimation (Figure \ref{CE}) and the \ac{AirComp} (Figure \ref{AirComp2}). 
The performance of the communications was also evaluated for a known channel to provide a bound. 

The performance of \ac{AirComp} was also evaluated for a known channel, as well as for known communication symbols (Figure \ref{AirComp2}). 

As can be observed from Figure \ref{BER}, the proposed algorithm closely approaches the bound where perfect \ac{CSI} is assumed, although the performance degrades with larger velocities due to the large Doppler shifts.
A similar trend can be observed for the channel estimation performance as seen from Figure \ref{CE}.

Finally, it can be observed from Figure \ref{AirComp2} that the \ac{NMSE} performance also matches the Genie-Aided case.

\vspace{-1ex}
\section{Conclusion}

We proposed a system for \ac{ICC} in time-varying \ac{mmWave} channels, which also keeps track of the channel variations.
The system uses a \ac{BiGaBP} algorithm aided by a channel prediction mechanism to estimate communication symbols and the channel time variations, while the target function estimate is computed after detection using the channel residual. 
A beamforming scheme is also proposed for the multi-user \ac{mmWave} communications based on the existing \ac{SVD} technique for point-to-point \ac{MIMO} communications to minimize interference in the channel, which eases detection.
The computing signal only uses a small fraction of the transmit power, which means that the communication task is not penalized by the integration of computing. 
In contrast, the computing performance is also almost completely independent of the receiver's knowledge of the communication symbols.

\bibliographystyle{IEEEtran}
\bibliography{references}

\end{document}